\documentclass[grl,draft]{agutex}
\usepackage{amssymb}
\usepackage[dvips]{graphicx}
\authorrunninghead{DE BARROS ET AL.}
\titlerunninghead{Source geometry of LP events at Mt Etna}

\authoraddr{C. J. Bean, School of Geological Sciences, University College Dublin, Belfield, Dublin 4, Ireland.}
\authoraddr{L. De Barros, School of Geological Sciences, University College Dublin, Belfield, Dublin 4, Ireland. (louis.debarros@ucd.ie)}
\authoraddr{I. Lokmer, School of Geological Sciences, University College Dublin, Belfield, Dublin 4, Ireland.}
\authoraddr{J.-P. M\'etaxian, LGIT, Universit\'e de Savoie-IRD-CNRS, 73376 Chamb\'ery, France. }
\authoraddr{G. S. O'Brien, School of Geological Sciences, University College Dublin, Belfield, Dublin 4, Ireland.}
\authoraddr{D. Patan\`e, INGV-Catania, Piazza Roma 2, 95125 Catania, Italia.}
\authoraddr{G. Saccorotti, INGV-Pisa, Via della Faggiola 32, 56126 Pisa, Italia.}
\authoraddr{L. Zuccarello, INGV-Catania, Piazza Roma 2, 95125 Catania, Italia.}

\begin{document}

\title{Source geometry from exceptionally high resolution Long Period event observations at Mt Etna during the 2008 eruption.}

\author{Louis De Barros\altaffilmark{1}, Christopher J. Bean\altaffilmark{1}, Ivan Lokmer\altaffilmark{1}, Gilberto Saccorotti\altaffilmark{2},  Luciano Zuccarello\altaffilmark{1,3}, Gareth S. O'Brien\altaffilmark{1}, Jean-Philippe M\'etaxian\altaffilmark{4}, Domenico Patan\`e\altaffilmark{3}}

\altaffiltext{1}{School of Geological Sciences, University College Dublin, Dublin, Ireland.}
\altaffiltext{2}{INGV-Pisa, Pisa, Italy.}
\altaffiltext{3}{INGV-Catania, Catania, Italy.}
\altaffiltext{4}{LGIT, Universit\'e de Savoie-IRD-CNRS, Chamb\'ery, France.}

\begin{abstract}
During the second half of June, 2008, 50 broadband seismic stations were deployed on Etna volcano in close proximity to the summit, allowing us to observe seismic activity with exceptionally high resolution. 129 Long Period events (LP) with dominant frequencies ranging between 0.3 and 1.2 Hz, were extracted from this dataset.  These events form two families of similar waveforms with different temporal distributions. Event locations are performed by cross-correlating signals for all pairs of stations in a two-step scheme. In the first step, the absolute location of the centre of the clusters was found. In the second step, all events are located using this position. The hypocentres are found at shallow depths (0 to 700 m deep) below the summit craters. The very high location resolution allows us to detect the temporal migration of the events along a dike-like structure and 2 pipe shaped bodies, yielding an unprecedented view of some elements of the shallow plumbing system at Mount Etna. These events do not seem to be a direct indicator of the ongoing lava flow or magma upwelling.
\end{abstract}


\begin{article}
 \section{Introduction}
Mt Etna is an active 3,330 m high stratovolcano located on the East coast of Sicily, Italy. An eruptive period began on the 10$^{th}$ of May 2008 with a powerful lava fountain in the South East Crater, one of the four main summit craters. An eruptive fracture opened on the 13$^{th}$ of May on the eastern flank of the volcano, in the ``Valle del Bove'' \citep[see e.g.][]{Napoli2008}. The eruption stopped on July 7$^{th}$ 2009. \\
Long Periods (LP) events, with frequencies ranging from 0.2 to 1.3 Hz on Mt Etna, are thought to be associated with resonance or transport of fluid in the volcano conduits  and are often considered as precursors to an eruption \citep{chouet96}. Locating these events can greatly improve our knowledge of the geometry of the plumbing system of the volcano. Furthermore, an accurate location can help us constrain moment tensor inversions leading to a better understanding of the source process.  \\
As LP signals have an emergent onset, classical travel-time inversion cannot usually be used to locate the source of these events. Several methods have been developed to locate them: semblance method \citep[e.g.][]{patane08}, array techniques with frequency-slowness analysis \citep{metaxian02}, amplitude decay  \citep{battaglia03}, coupled inversion for location and moment tensor \citep{kumagai02} or travel time inversion with improved pick readings achieved through stacking similar events \citep{saccorotti07}.\\
In the past, several studies have been conducted on LP events from Mt Etna \citep{falsaperla02,saccorotti07,lokmer07,patane08}. They found LP sources located below the summit area at shallow depths, i.e. 0-2000 m. No rapid changes in the LP source locations were detected, only small changes in the LP event characteristics before and after eruptive periods. The link between LP events and eruptions is still an open question \citep{patane08}. The location of tremor sources show two connected dike bodies oriented in a NW-SE direction extending from sea level to the surface \citep{patane08}, which is in agreement with geodetic data \citep{Bonaccorso2002} and LP source mechanisms \citep{lokmer07b}. However, locations of LP events have not yet shown any clear structural geometry on Mt Etna. \\

The aim of this paper is to obtain information about LP source distribution. We use the observations from a temporary deployment with 50 broadband stations. LP events were extracted and classified from this dataset. We then focused on the location of these events using the time delays between closely spaced stations measured by cross-correlation. The resulting high resolution source locations show outstanding well-defined geometries with an unprecedented short term temporal variation.

\section{Data}

A total of 50 stations with three component broadband sensors (30, 40 or 60 s cut-off period with 5 or 10 ms sampling rate) were deployed on Etna volcano between the 18$^{th}$ of June 2008 and the 3$^{rd}$ of July 2008. This included 16 permanent stations from INGV, Italy and 34 temporary stations from University College Dublin (Ireland), Universit\'e de Savoie (France) and INGV (Italy). Such a large number of stations is quite unusual on a volcano. In particular, 30 stations were located less than 2 km from the summit area (see fig. 1). \\
Before analysing the data, we deconvolve the instrument response from the recorded signals.  To extract the Long Period events, we use a STA/LTA  method (2s over 20s window lengths, with a threshold of 2.5) on the bandpass filtered data (0.2-1.5 Hz), which give us approximately 500 events. We then classify these events using a cross-correlation analysis between all pairs of signals \citep{saccorotti07}. We keep the events that give a correlation coefficient greater than 0.9 with all events on at least 3 out of the 4 permanent stations close to the summit. We obtain two different families with a similar number of events (63 and 66, resp.). \\
Figure 2a) shows the temporal distribution of these events. The first family is only present in the first two days of the experiment (18$^{th}$-19$^{th}$ of June), while the second family is distributed over the first four days. After June 22$^{nd}$, the amplitudes of the LP events decrease by an order of magnitude. In the same period, the tremor amplitude increases. Since both the LPs and tremors are in the same spectral range, it makes it difficult to recognize and extract additional LP events after June 22$^{nd}$.\\
The waveforms and the spectral content of the stacked events for both families are shown in figure 2b). Though the waveforms are quite similar, the spectral peaks are not the same for both families. The second one has a sharper spectrum, with a peak frequency slightly higher than family 1. The waveform similarity within each group suggests spatially close sources with a similar mechanism, while the source position and/or the mechanism have to be different between the two families.

\section{Method}
The location of the LP events is computed in two steps. We first find the mean position for each family and we then locate the individual events using this first position. As the LPs are emergent (as seen in fig. 2b) ), it is impossible to directly measure the arrival time. Instead, we choose to use cross-correlation between stations $i$ and $j$ ($i\neq j$) to obtain the time delays $t^{obs}_{ij}$. \\
In the first step, we improve the Signal to Noise Ratio by stacking similar events. For a hypothesized source position $X_s(x_s,y_s,z_s)$, we compute the distance between the source and each station. The propagation medium is assumed to be homogeneous which leads to the approximation of spherical wavefronts as the  source-to-receiver distances are short. Theoretical time delays $t^{th}_{ij}(X_s)$  between pairs of stations are then obtained by dividing the distance difference by the wave velocity.
We then use a grid search to find the position $X_s$ which minimizes the misfit function defined by:

\begin{eqnarray}\label{eq1}
R(X_s)=\sum_i \sum_{j\neq i}W_{ij}C_{ij}\left(t^{obs}_{ij}-t^{th}_{ij}(X_s)\right)^2~,
\end{eqnarray}

where $C_{ij}$ is a weight related to the correlation coefficient  $c_{ij}$ ($C_{ij}=[\xi_c/(1-c_{ij})]^{2}$), and $W_{ij}$ a correction factor inversely proportionnal to the time delays between stations ($W_{ij}=exp[-{t^{obs}_{ij}}^2/\xi_w$]), as we consider that errors increase with the propagation distance. $\xi_c$ and $\xi_w$ are normalizing constant which can be adjusted.\\
In this first step, we found the mean location $X_0$ of the hypocentre for each family. However because individual events are almost monochromatic and quite noisy, one or several cycles can be accidently skipped during the cross-correlation procedure, giving incorrect time delays. To avoid this problem and to refine the grid search, we introduce a second step, locating the individual events in each family using the mean positions of that family as reference positions. We compute the theoretical time delays $t^{th}_{ij}(X_0)$  between each pair of stations using the source position found in the first step. Events are shifted according to this theoretical time, and by cross-correlation we measure the residual time $\delta t^{obs}_{ij}=t^{obs}_{ij}-t^{th}_{ij}(X_0)$. As the source positions are close to each other, this time is much smaller than the central period of the events. The location is then found by minimizing the sum of the squared differences between observed and theoretical residual times for a grid of source position. This involves solving equation \ref{eq1} in the same manner as in step 1.\\

Although this location method does account for the topography, it does not take the wave propagation effects (free surface, velocity heterogeneities) into account. However, this is balanced by the large number of stations and their close proximity to the source. Synthetics tests are performed to check the accuracy of the location (see figure 4, auxiliary materials). Errors are found to be very small in the horizontal plane ($<$10m) and bigger in the vertical direction ($<$100m).\\

\section{Location results}
We locate the events for both families with the method from section 3. We only use stations close to the summit, i.e. 25 stations for family 2 and 19 for family 1, according to the available data. We compute the stacked events for both clusters and we use 2x2x2 km grid with a spacing of 50 m. In order to determine the velocity, we run inversions for wave velocities between 1.2 and 3.2 km/s in steps of 0.2 km/s. The lowest residuals are obtained for a velocity of 1.8 km/s, which is in agreement with near-surface velocity measurements on the Mt Etna \citep{patane08}. The hypocentre positions are found to be at $X_0=(499.4,4178.7,2.95)$ km for family 1 and $X_0=(499.5,4178.45,3)$ km for family 2. These positions are then used as the reference for the second step. Again we search for the optimum velocity, yielding a value of 1.8 km/s for the shallowest events and 2.2 km/s for the deepest ones. We search for the source positions in a 600x600x900 m grid with a 10 m spacing. The results of this location procedure are shown in figure 3.  \\

Firstly, the events for both families are shallow: from 0 to 800 m for family 1 and from 0 to 400 m for family 2. The epicenters of the events are close to the summit craters, and are slightly different for both families. The colourscale in figure 3 and in the animation in auxiliary material indicates the origin times of the events. We see that the source positions of family 1 events migrate upwards (but not systematically) from depth to the surface between the 18$^{th}$ and the beginning of the 19$^{th}$ of June. During this period, events of family 2 are located in the area occupied by the shallowest events of family 1 and move downwards from the surface to a depth of 400 m. After the middle of June 19$^{th}$, the hypocentres of family 2 migrate back towards the surface using another path. \\

To quantify the geometry of the clusters revealed by the hypocentres position, we use a simplified Principal Parameter method \citep{michelini86}. The eigenvectors and eigenvalues of the spread matrix (i.e. covariance matrix of the hypocenters) give us the directions and lengths ($L_1>L_2>L_3$) of the three principal parameter axes (see table 1) which allow us to determine the best ellipsoid which fits the source positions. As the cluster geometries are clearly defined, we apply this analysis to the whole family 1 (F1) and  to the two well defined branches (called F2a and F2b in table 1) of the family 2.  The azimuths of the principal axes (noted $\phi_1$ and $\phi_3$ for major and minor axes) are measured from north clockwise and corresponding inclination angles $\theta_1$ and $\theta_3$ are from the horizontal plane, positive downward.\\

Following \cite{michelini86}, a planar structure can be defined with $L_1/L_3$ $\geq$ 2.5 and $L_2/L_3$ $\geq$ 1.75. A pipe shape structure will have $L_1/L_3$ $\geq$ 2.5 and $L_1/L_3$ $<$ 1.75. This suggests (see tab. 1)  that the cluster geometry of family 1 is more dike-like, while the two branches of family 2 are closer to pipe shapes. Family 1 mainly shows a subvertical planar geometry with normal defined by $\phi_3$=31\deg and $\theta_3$=5\deg.   The two clusters of family 2 are elongated in directions $\phi_1$=74\deg, $\theta_1$=26\deg for F2a and $\phi_1$=8\deg, $\theta_1$=46\deg for F2b. However, the two clusters F2a and F2b can be merged in a single cluster as they belong to a same plane whose normal is defined by $\phi_3$=137\deg and $\theta_3$=37\deg.  This can be interpreted as the presence of a planar structure for family 2 within which the LP source locations move, branching into two directions.

\begin{table}[p]
\caption{Main characteristics of the structures: azimuth ($\phi_1$ and $\phi_3$) and inclination angle ($\theta_1$ and $\theta_3$) of the major and minor principal parameter axes; Ratio between major ($L_1$), intermediate ($L_2$) and minor ($L_3$) principal parameter axes; strike and dip (=azimuth and plunge for the pipe-shape F2a and F2b).}\label{tab1}
\begin{tabular}{|c|c|c|c|c|c|c|c|c|}
\hline
 Fam. & $\phi_1$ (\deg) & $\theta_1$ (\deg) & $\phi_3$ (\deg) & $\theta_3$ (\deg) & $L_1/L_3$ &$L_2/L_3$ & strike (\deg) &dip (\deg)\\  
\hline
 F1&  306 & 46 & 31 & 5 & 5.2 & 2.5 & N301 & 85E\\
F2a &  254 & 26& 147 &  42& 4.8 & 1.5 & N254 & 26W\\
 F2b & 8 & 46 & 172 &  44 & 4 & 1.3& N8 & 46N\\ 
\hline
\end{tabular}

\end{table}

\section{Discussion and Conclusion}
Two families of LP events (63 and 66 events selected) were found in the first four days of an experiment carried out between the 18th of June and the 3rd of July, 2008. The data were recorded by an exceptionally high-resolution network, consisting of 50 broadband stations deployed in the close proximity of the source ($<$ 2 km), thus enabling us to locate the source positions with a very high precision. The location of the stacked events and of all the individual events were determined and are on average in broad agreement with previous studies \citep{saccorotti07,patane08}. However, LP event distributions show outstanding well-defined geometry with an unprecedented temporal evolution. Hypocentres moved, in a 96 hour period, from a depth of 800 m to the surface through a planar structure (family 1 events) which branches at 300 m below the summit craters into two structures which have more pipe geometry (family 2 events). The deeper structure is subvertical and striking NW-SE (N301\deg, 85\deg E) in agreement with the results of \citet{patane08} and \citet{lokmer07b}, while the two structures of family 2 are aligned in a SW-NE striking plane (N47\deg, 53\deg W). As some events share similar positions but belong to different families, it suggests that the difference between the seismograms comprising the two families is due to the source mechanism.\\
Mt Etna volcano was active during the experiment with lava flowing from an eruptive fissure on the eastern flank of the volcano. The highest part of the active vents were 500 m below and 1 km from the summit craters. There is no visible evidence of a change in terms of eruptive output associated with the LP migration nor with the energy decrease of the LP events after the 22$^{nd}$ of June. It is clear that these LP events are not representative of the whole eruption, but our results show that the LP activity recorded at this time does not seem to be an indicator of the ongoing flank lava flow nor of magma upwelling. Moreover, the LP source area and their disappearance after the 22$^{nd}$ of June can be related  to the family 2 events of \cite{patane08} which were recorded only for a few weeks after the lava fountains of 2007. This suggests that the events found in this study are more likely the end of the response to the lava fountain of the 10$^{th}$ of May 2008. One possible hypothesis is that these seismic events are associated with magma trapped in plugged conduits leading to the summit craters. Another hypothesis may be that LP events are not directly related to magma, but rather to gas, which is continuously emitted from the summit craters. Determining the source mechanisms by moment tensor inversion \citep{kumagai02,lokmer07b} will provide more insights into the process generating these events. 

\begin{acknowledgments}
L. De B. and G.S. O'B. were part funded by the department of Communications, Energy and Natural Resources (Ireland) under the National Geosciences programme 2007-2013. Financial assistance for field work from University College Dublin and the EU Volume project is acknowledged. A particular thanks to  M. M\"ollhoff, J. Grangeon, P. Bascou,  M. La Rocca, D. Galluzzo and S. Rapisarda for assistance in the field experiment.
\end{acknowledgments}

\begin{figure}[p]
\includegraphics[height=7cm]{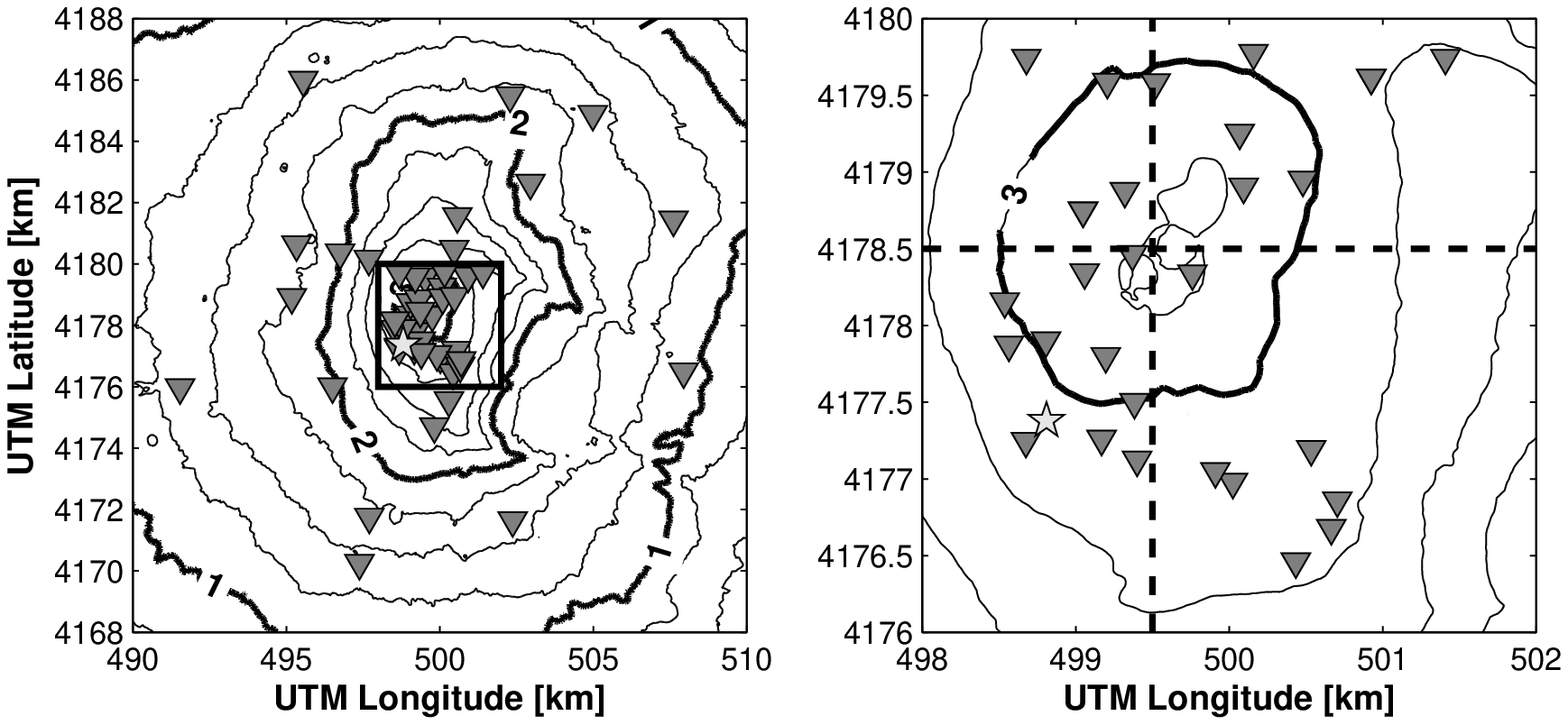}
\caption{Broadband stations positions. Left: All stations deployed on the volcano, Right: Stations within 2 km of the summit used in this study. Station ECPN is marked by a star. Elevation contour step is 250m. Thick dotted lines in the right panel indicate the positions of the cross-sections used in this study.} \label{fig1}
\end{figure}

 \begin{figure}[p]
    \includegraphics[height=8cm]{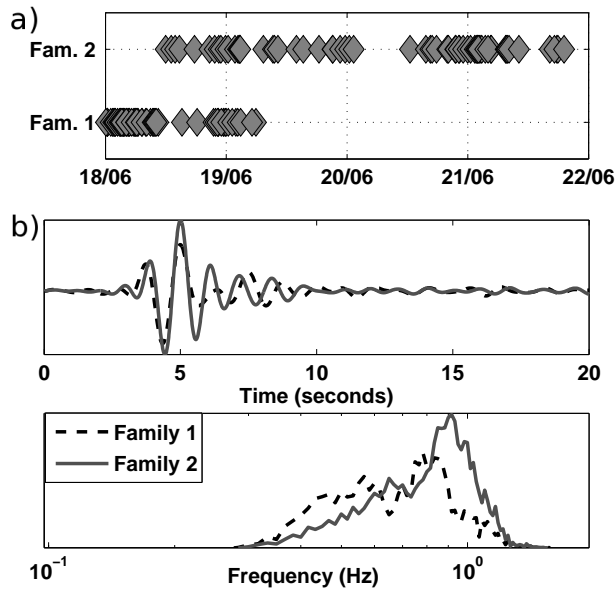}
 \caption{a) Temporal distribution of the LP events. Families 1 and 2 contain 63 and 66 events, respectively. b) Waveform and spectral content for stacked events (filtered between 0.2 and 1.5 Hz) at station ECPN (see fig. \ref{fig1}) for both families (fam 1=dashed line, fam 2=solid line).}\label{fig2}
 \end{figure}

\begin{figure}[p]
\hspace{-1cm}\includegraphics[height=10cm]{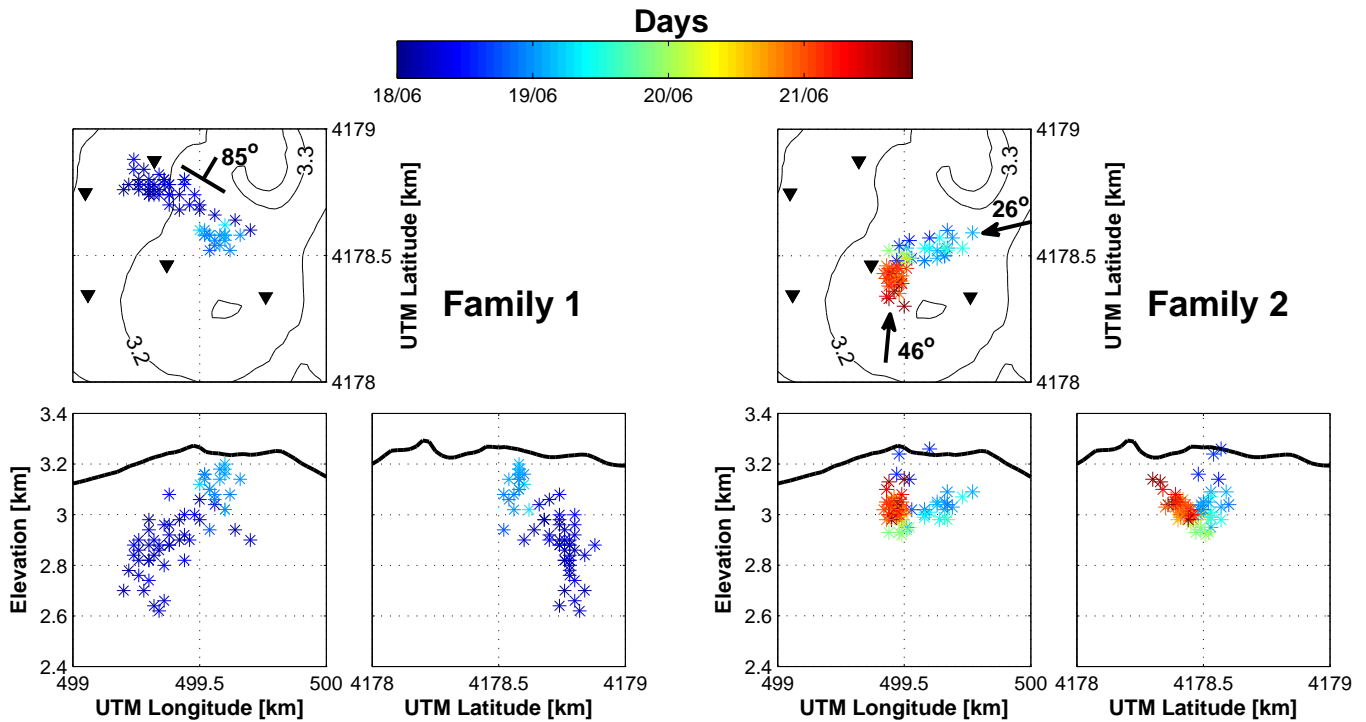}
\caption{Location of all the LPs for family 1 (left) and 2 (right) with colours indicating temporal evolution. Colourscale (days) is common for both family. Views are from above, South and West. Triangles are some of the broadband stations. The side views correspond to the crosssections indicated in fig. \ref{fig1}.``$T$'' and ``$\rightarrow$'' symbols represent the planar and pipe-shape structures, with dip and plunge angles respectively.}\label{fig3}
\end{figure}

\end{article}
\end{document}